# Growth and structural characterization of pyramidal site-controlled quantum dots with high uniformity and spectral purity

**Valeria Dimastrodonato** [*,1], **Lorenzo O. Mereni**[1], **Robert J. Young**[1] **and Emanuele Pelucchi**[1]

[1] Tyndall National Institute, University College Cork, "Lee Maltings", Cork, Ireland

* Corresponding author: e-mail valeria.dimas@tyndall.ie, Phone: +353 21 420 4195, Fax: +353 21 4270271

**Abstract**

This work presents some fundamental features of pyramidal site-controlled InGaAs Quantum Dots (QDs) grown by MetalOrganic Vapour Phase Epitaxy on patterned GaAs (111)B substrate. The dots self-form inside pyramidal recesses patterned on the wafer via pre-growth processing. The major advantage of this growth technique is the control it provides over the dot nucleation position and the dimensions of the confined structures onto the substrate.

The fundamental steps of substrate patterning and the QD formation mechanism are described together with a discussion of the structural particulars. The post-growth processes, including surface etching and substrate removal, which are required to facilitate optical characterization, are discussed. With this approach extremely high uniformity and record spectral purity are both achieved.

**1 Introduction** Semiconductor QDs are subject of a wide variety of studies ranging from fundamental physics, quantum electrodynamics [1] to quantum information and computing [2]. Despite the gamut of potential applications, some practical questions are still open. A precise control over the position and dimensions of the dots and a versatile tuning mechanism of their electro-optical properties are desirable. The most commonly used growth technique, Stranski-Krastanov (SK), is based on the spontaneous nucleation of the confined nanostructures on the substrate and it is an intrinsically random process, giving rise to QDs ensembles lacking in order, typically with a large inhomogeneous distribution of physical properties. "Site-control" growth techniques are an excellent alternative, but in most cases they offer diminished optical properties, due to unintentional incorporation of contaminants following the pre-growth processing of the substrate. It has been demonstrated that site-controlled (In)GaAs QDs embedded between AlGaAs barriers can be grown successfully on GaAs (111)B substrates [3]. Unlike other site-controlled methods [4], this approach offers both high uniform properties and a quality on-par with traditional self-organised methods [5]. These results are facilitated by the self-formation mechanism of the dot buried in the epitaxial structure, and the excellent layer and interface qualities it affords.

An evolution of this pyramidal structure is discussed here: primarily the AlGaAs barriers are substituted with GaAs. We have previously reported that with this substitution and careful reactor handling and sources purification the optical properties of the excitons confined in systems are drastically improved [6].

We present here relevant details, not previously published, on the sample processing and structural characterization. A brief description of the pre-growth substrate patterning is followed by a discussion of the formation process and a detailed study of the pyramidal structure. We also describe the post-growth processes, chemical surface etching and substrate removal, essential to prepare the QD systems for optical characterization.

**2 Growth and Structural characterization** All samples were grown in a horizontal Aixtron 200 reactor with purified $AsH_3$ and TMGa, TMAl, TMIn as group V and III sources respectively. Before each growth the GaAs (111)B substrate was processed via photolithography and a chemical selective wet etch (in a ~1% $Br_2$:Methanol solu-tion) to create a uniform dense hexagonal lattice of 7.5μm pitch pyramids, acting as nucleation seeds during growth. As result of the patterning process, each pyramid exposes lateral facets with crystallographic orientation (111)A and very sharp edges and tip (Fig. 1).

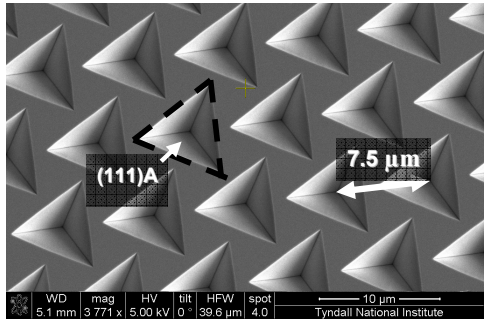

**Figure 1** Top view Scanning Electron Microscopy (SEM) image of a GaAs (111)B substrate after the patterning process. Orientation of the sidewalls and pyramidal pitch are labelled. The high uniformity and the sharp edges and tips are evident.

The typical layer structure for a quantum dot grown in a pyramidal recess by MetalOrganic Vapour Phase Epitaxy (MOVPE) consists of: a thick GaAs buffer, which traps the unintentional contaminations incorporated during the patterning steps; an $Al_{0.8}Ga_{0.2}As$ protection layer for the post-growth back etching process (performed to improve the light extraction from the dot); outer $Al_{0.55}Ga_{0.45}As$ barriers, helping to better confine the carriers around the dot; inner GaAs barriers which embed the $In_xGa_{1-x}As$ dot layer. To avoid the oxidation of the top surface, due to the presence of Al, the growth is capped with a thin layer of GaAs. Extremely high V/III ratios, in a low pressure reactor environment, and growth temperatures ranging from 730°C to 800°C (thermocouple reading) are necessary to reduce the carbon incorporation and guarantee a good interface quality. Moreover a low growth rate, nominally equal to 0.5μm/h, was chosen to avoid the unwanted formation of crystal defects.

During the growth, as well known from previous studies [7], the triangular open face of the recesses becomes hexagonal-shaped due to a slight change in the as-grown sidewalls from the as-etched profile. A different QD system ending with the only GaAs inner barrier was grown in order to study this faceting effect. Fig. 2 shows a top view of such a structure observed by SEM: the as etched (111)A walls evolve toward vicinal (111)A facets, thus changing the triangular face into a hexagon. The planarization process induced by the recess filling, on the other hand, develops a very rich faceted structure, significantly different from similar observations in the AlGaAs system [7].

Fig. 3 is an Atomic Force Microscopy (AFM) image of a typical pyramidal QD system profiled in cross section. The measurement in such geometry is performed after cleaving through the substrate along the (110) direction (perpendicular to one of the triangular sides). If the cleave position on the wafer is close to the central axis of the pyramids, the dot can be observed very clearly [7] (see also inset in figure 3 for a pictorial explanation).

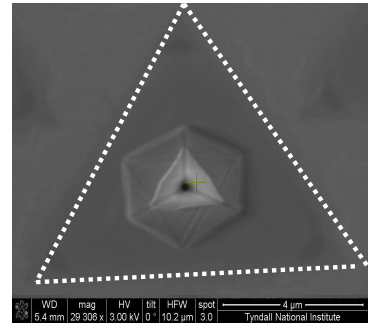

**Figure 2** A top-view SEM image of an as-grown pyramidal QD structure. The upper GaAs layer (the top AlGaAs layer was intentionally avoided for this sample) shows the hexagonal evolution from the original triangular shape. The total thickness of the structure is such that the layers significantly filled the original recess, approaching (but not reaching) what in the "pyramidal dot" community is known as the "closing threshold": i.e. when the pyramidal recesses start to be wildly filled by the growth process and no ordered crystallographic profile can be maintained.

From Fig.3 it is actually possible to examine the mechanisms responsible for dot formation. Since a QD is a nanostructure confined in three dimensions, its successful formation critically relies on a good lateral confinement, provided by a sharp bottom of the pyramid. As shown previously in Fig. 1, the selective chemical etching step delivers a near-perfect V-shaped tip, but during the epitaxial growth an evolution toward a self-limited profile with a wider base occurs. The width of the bottom is a result of a competition between growth rate anisotropy and capillarity effects [8] and therefore it strongly depends on the diffusion properties of the adatoms. In this context the employment of GaAs as an inner barrier material instead of Al-GaAs, which was successfully utilized for similar pyramidal structures in the past [3,5,7], becomes critical. The high mobility of Ga adatoms in fact enriches capillarity effects resulting in a widening of the self-limited profile base, which could compromise a good lateral confinement. Growth parameters may then need to be optimized. Previous studies by Biasiol et al. [9] for example, concerning the formation of Quantum Wires (QWRs) in V-grooved GaAs substrates, demonstrated the necessity of significantly decreasing the growth temperature (from ~700°C to even as low as ~550°C) in order to obtain a relatively sharp bottom profile. It must be noted that in a V-groove structure a further widening of the self-limited base is caused by the presence of (311)A oriented lateral facets which separate the (100) bottom and the (111)A sidewalls.

For a pyramidal template the (311) high index planes don't form and a resulting narrower self-limited profile can be achieved even at relatively high temperature (~730˚C). As indicated in Fig.3 (visible and measurable at the interface between layer 4 and 3, top of the image) a GaAs base of ~67 nm was reached and the correspondent lateral confinement allowed the formation of the dot, indicated in the same figure. It is noteworthy that this is the first structural (AFM) evidence of the Pyramidal InGaAs QD formation in GaAs barriers reported in the literature. The same mechanism responsible for the dot evolution makes the formation of Lateral Quantum Wires (LQWRs) possible, which are present in the image, and Lateral Quantum Wells (LQWs). These form respectively along the edges and the sidewalls of the structure [7].

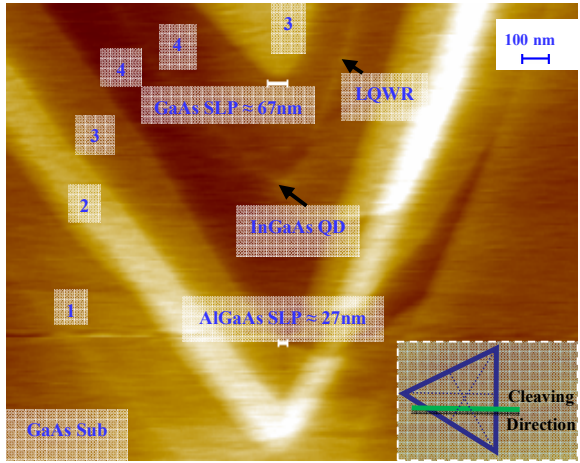

**Figure 3** An AFM image taken from a quantum dot structure containing a 1.5nm thick $In_{0.35}Ga_{0.65}As$/GaAs QD in cross-section. The layer sequence is: GaAs buffer (1), $Al_{0.8}Ga_{0.2}As$ etch-stop layer (2), outer $Al_{0.55}Ga_{0.45}As$ barriers (3), inner GaAs barriers (4), $In_{0.35}Ga_{0.65}As$ QD (labelled). The self-limited profile (SLP) bases were measured both for GaAs and AlGaAs. The quantum dot itself is clearly visible, this is due to the strain (caused by the lattice constant difference between GaAs and InGaAs) which relaxes once the substrate is cleaved. The contrast between AlGaAs and GaAs is a result of the growth of a thin oxide layer upon AlGaAs, resulting in a different height being measured for the two materials by the AFM tip [7].

**3 Preparation of the samples for optical characterization** Following the growth, the samples were prepared for µ-PhotoLuminescence (µPL) measurements, to test their optical properties. Between the planar substrate (111)B and the sidewalls (111)A, some irregular facets tend to form. Their unwanted emission can obfuscate the signal emitted from the dot. To prevent this they are removed with a simple "surface-etching" process [10]. A positive photoresist S1805 is deposited at 5000 rpm for 30 seconds on the as-grown sample. After an oxygen plasma etch (at 75 W with a chamber pressure of 0.8 Torr), only the upper facets of the structure are freed from the resist and the remaining layer protects the central part of the pyramid in the following chemical etching step. This allows to remove the irregular sidewalls in a solution of $H_2SO_4:H_2O_2:H_2O=1:8:160$ for 70 seconds. Any remaining resist is then removed in an ultrasound bath of acetone for 5 minutes, this is repeated twice. Finally the sample is cleaned in an isopropyl bath. To improve the light extraction from the pyramidal structure a further post-growth process known as "back-etching" is performed. This involves removing the substrate [10], which is achieved with an all-chemical procedure. The sample is coated with a ~200 nm thick gold layer to give good mechanical support. It is then attached onto a second substrate, upside-down with a black wax. A primary chemical etch in ammonia solution diluted in $H_2O_2$ removes the bulk of the substrate. The dilution in $H_2O_2$ acts to change the PH of the solution adjusting the etching rate. First a PH of 8.7 allows a relatively fast etching of GaAs substrate (~100 nm/hour). The PH is then reduced to 8 to better control the etching speed. A secondary chemical etch consisting of a solution of $C_6H_8O_7:H_2O:H_2O_2=3:3:1$ is used to uniformly etch the exposed pyramids. This last step is highly beneficial in preventing damage to the pyramids. At the end of the process the pyramidal structures are orientated apex-up (Fig. 4). In this geometry the total internal reflection from the lateral facets of the pyramid are avoided and the tip acts to guide the light from the semiconductor structure. The result is a remarkable increase of the emitted light.

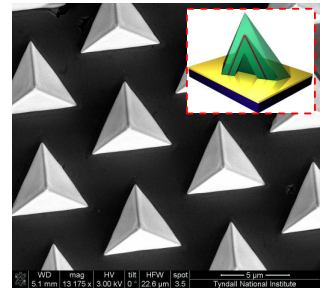

**Figure 4** Top view SEM image of "back-etched" pyramids. The apex is now free standing. In the inset a cross-section of an up standing pyramid is illustrated.

For µPL characterization the samples are mounted into a closed-cycle helium cryostat and cooled to ~10K. Typically, they are excited under non-resonant conditions with a laser emitting at 658nm. The self-formation mechanism employed here, which distinguishes these samples from other site-control based QDs [6], allows the high spectral purity to be preserved. Extremely narrow Full Width at Half Maximum of the emitted exciton peaks were measured and the best sample (a ~0.5 nm thick $In_{0.25}Ga_{0.75}As$/GaAs QD system) exhibits linewidths of 18µeV-20 µeV [6]. Record uniformity of the growth process and optical properties were also reached: an inhomogeneous broadening of 2.3meV was measured [6].

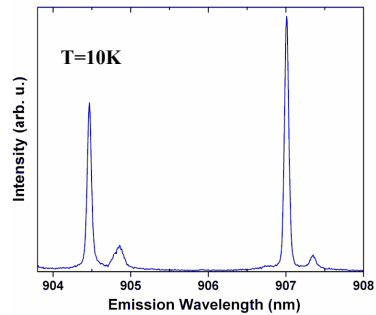

**Figure 5** A low temperature μPL spectrum of exciton peaks emitted from two coupled pyramidal $In_{0.25}Ga_{0.75}As$/GaAs QDs vertically stacked and separated by a 2nm barrier.

Pyramidal QD systems are advantageous over other site-controlled schemes because of their versatility. Their position/dimensions can be easily tuned for achieving for example good coupling to microcavities and/or a good tailoring of their optical properties [11]. The emission energy can be tuned by changing the thickness and/or the composition (In concentration) of the dot layer. Vertically stacked dots can also be coupled in a pyramidal system and engineering of coupling properties can be achieved by controlling the exciton wavefunction overlap, by either modifying the thickness of the barrier between the coupled dots or electrically driving the dot system. Fig. 5 shows some preliminary measurements of an emission spectrum of the excitons confined in two "coupled" $In_{0.25}Ga_{0.75}As$/GaAs dots separated by a nominal GaAs barrier of 2nm, showing an emission which is significantly red-shifted if compared to isolated, "non-coupled" structures [6]. More comprehensive work regarding the optical properties of such coupled structures is in progress, and will be presented elsewhere.

**4 Conclusions** Structural characteristics peculiar of pyramidal site-controlled QDs have been described. Details on the growth parameters and principles of the dot formation have been discussed. We also have presented the surface and back etching post-growth processes which allow us to prepare the samples for μPL measurements. Record uniform optical properties and very narrow excitonic peaks can be achieved with the approach showed here. The physical properties of the dot system can be easily tuned by altering the growth parameters thus engineering its optical properties, making this dot system highly versatile.

**Acknowledgements** This research was enabled by the Irish Higher Education Authority Program for Research in Third level Instruction (2007-2011) via the Inspire programme and by Science Foundation Ireland under grant 05.IN.1/I25. We are grateful to Dr. Kevin Thomas for the support with the MOVPE system.